\documentclass[preprint,showpacs,showkeys,preprintnumbers,amsmath,aps,amssymb,superscriptaddress]{revtex4}
\usepackage[english]{babel}
\usepackage{graphicx}
\usepackage{dcolumn}
\usepackage{bm}
\usepackage{hyperref}
\usepackage{color}
\usepackage{lineno}
\usepackage{amssymb,amsbsy} 
\usepackage{amsthm}
\usepackage{amsmath}
\usepackage{verbatim}
\usepackage{subfigure}
\usepackage{soul}
\usepackage{ulem}
\usepackage{slashed}
\usepackage{rotating}
\usepackage{textcomp}
\usepackage{booktabs}
\usepackage[all]{xy}
\usepackage{cleveref}
\usepackage{ragged2e}

\begin{document}

\title{ The square-well fluid: a thermodynamic geometric view}

\author{J. L. L\'opez-Pic\'on}
\email{jl\_lopez@fisica.ugto.mx}
\affiliation{Divisi\'on de Ciencias e Ingenier\'ias Campus Le\'on,
    Universidad de Guanajuato, A.P. E-143, C.P. 37150, Le\'on, Guanajuato, M\'exico.}
\author{L. F. Escamilla-Herrera} \email{lenin.escamilla@correo.nucleares.unam.mx}
\affiliation{Divisi\'on de Ciencias e Ingenier\'ias Campus Le\'on,
    Universidad de Guanajuato, A.P. E-143, C.P. 37150, Le\'on, Guanajuato, M\'exico.}
\affiliation{Instituto de Ciencias Nucleares, Universidad Nacional 
	Aut\'{o}noma de M\'{e}xico,\\ Mexico City 04510, M\'{e}xico.}
\author{Jos\'e Torres-Arenas} \email{jtorres@fisica.ugto.mx}
\affiliation{Divisi\'on de Ciencias e Ingenier\'ias Campus Le\'on,
    Universidad de Guanajuato, A.P. E-143, C.P. 37150, Le\'on, Guanajuato, M\'exico.}

\date{\today}

\begin{abstract}
\noindent The square-well fluid with hard-sphere diameters is studied within the framework of Thermodynamic Geometry (TG). Coexistence 
and spinodal curves, as well as the Widom line for ranges $\lambda^{*} = 1.25, 1.5, 2.0, 3.0$ for this fluid are carefully studied 
using  geometric methods. We are able to show that, unlike coexistence curves, an exact result can be given along all the thermodynamic 
space and for all potential ranges for spinodal curves. Additionally, $R$-Widom line which is defined as the locus of extrema of the 
curvature scalar is calculated as a function of the potential range, satisfying near the critical point a kind of Clausius-Clapeyron equation. 
It is argued that this relation could be a signature of certain type of characteristic phase transition when crossing the boundary of 
the Widom line. 
\end{abstract}
 
 \keywords{Statistical Mechanics, Square-well fluid, Thermodynamic Geometry}
\pacs{05.20.Jj, 95.30.Sf, 95.30.Tg}

\maketitle

\section{Introduction}

One of the simplest fluid models, which has been widely used in the literature is the Square-Well (SW) fluid. This potential  includes both, attractive and repulsive interactions. The SW potential, $\phi(r)$, for any pair of particles at a distance $r$ apart is defined by the following functional relation,
\begin{equation} \label{eq:1}
	\phi (r) =
	\begin{cases}
	\infty \,, & r< \sigma \,, \\
	-\epsilon \,, & \sigma \le r \le \lambda \sigma \,, \\
	0 \,, & r > \lambda\sigma \,,  \\
	 \end{cases}
\end{equation}
where $\sigma$ is the hard-sphere diameter for each particle, $\epsilon$ is the potential depth and $\lambda$ is the range of the SW
interaction. This model has been extensively used to study structure and thermodynamic properties of fluids from different perspectives. 
For instance, several integral equation theories have been numerically solved for different ranges of the SW potential, including the
Percus-Yevick, hypernetted chain and mean-spherical approximations \cite{Tago1973,Henderson,Henderson1976}. Monte Carlo simulation 
has been carried out for SW potential since the seminal works of Alder et al. \cite{Alder1972} and many other simulation works 
have been done for  this interaction  considering different potential ranges, particularly, for $\lambda^* \equiv \lambda
\sigma^{-1} = 1.5$, due to the close resemblance of its thermodynamic properties with those of the Lennard-Jones fluid model.

Among all the theoretical studies carried out for the SW potential available in literature, theoretical and analytical equations of 
state have an outstanding place. There are several  proposed theoretical equations of state for the SW fluid which describe the behavior 
of this model with more or less accuracy depending on the potential range or the region of its corresponding thermodynamic space 
(due to the fact that it is always difficult to accurately describe the region near the critical point)
\cite{Benavides1999,Vega1992,Khanpour2011,DelRio2002,Patel2005,Martinez-Ruiz2017,Zeron2018,Sastre2020,Akhouri2020,DeSantiago2021,Sandoval-Puentes2022}. \\
It is remarkable that despite the large number of works devoted to study and explore the SW model, to our knowledge, there is not a comprehensive study for this system from the perspective of thermodynamic geometry. 

On the other hand, different approaches to introduce differential geometry on thermodynamics were developed in the past years; for instance, 
Hessian metrics for thermodynamics were postulated by Weinhold \cite{Weinhold1975} and Ruppeiner \cite{Ruppeiner1979,Ruppeiner1995}; 
in both approaches a metric is defined in the thermodynamic equilibrium space via the Hessian of a given thermodynamic potential. 
The Weinhold metric is constructed with the Hessian of internal energy, while Ruppeiner metric is given by considering the negative 
Hessian of entropy. It was later shown that both metrics are related by a conformal factor \cite{Salamon1984}.\\ 
Theoretical framework of these Hessian metrics is commonly known in the literature as Thermodynamic Geometry. Since their inception during the 1970's there have been a significant number of works devoted to study the interesting physical implications of these structures in Thermodynamics, particularly for Ruppeiner metric \cite{Ruppeiner1979,Salamon1983,Salamon1984,Janyszek1989,Crooks2007}. Additionally, connection between these metrics and statistical inference theory and statistical mechanics have been also explored \cite{Amari1982,Mrugala1990,Brody1995,Amari2009}. More recently, a new geometric description of thermodynamics have been proposed \cite{Quevedo2007}, extending the idea of invariance of thermodynamic potentials under Legendre transformation into a geometrical language. In that sense, it is possible to describe Legendre transformations in the thermodynamic phase space (Hermann introduced the concept of a phase space for thermodynamic systems as a  manifold of $(2n+1)$ dimensions equipped with a contact structure, whose maximally integrable submanifolds constitute the thermodynamic equilibrium space \cite{Hermann1973}), as a set of transformations which leave the Gibbs fundamental one-form invariant. This is the idea that inspire the formulation of Geometrothermodynamics, whose main premise is that invariance under Legendre transformations must be satisfied by all geometric structures. In this formalism, a metric on thermodynamic equilibrium space emerges as a consequence of imposing this condition obtaining the most general solution for an invariant metric under this set of transformations \cite{Quevedo2022}.

In previous works, thermodynamic geometry has been applied to study the thermodynamic behavior of a variety of pairwise additive interaction potentials modeling the fluid phase, as an example, May and Mausbach \cite{May2013} studied the behavior of thermodynamic response functions and the scalar curvature in the supercritical region for the Lennard-Jones fluid. More recently, Jaramillo-Gutierrez et. al \cite{Jaramillo2019,Jaramillo2020}, applied the $R$-crossing method to reproduce the coexistence curves of Yukawa and Mie fluids in the vicinity of the critical point, finding a dependence of the method with the potential range. Additionally, in \cite{Jaramillo2022}, the Widom line (i.e., a line separating a gas-like from a liquid-like phases in the supercritical region) was studied for the a family of Yukawa and Mie potentials, more specifically, the so called $R$-Widom line, defined by the locus of extrema for the isotherms  of the scalar curvature in the supercritical region. The main motivation of this work is to provide a study of the SW fluid under theoretical framework of thermodynamic geometry. Scalar curvature as a function of the SW potential range will be obtained; $R$-Widom lines for this family of potentials will be studied and related to a Clausius-Clapeyron equation; $R$-crossing method will be applied in order to calculate coexistence curves for different ranges of the SW fluid; finally, spinodal curves for the selected potential ranges will be provided, emphasizing that, unlike the coexistence curves, spinodal ones, can be fully constructed via thermodynamic geometry.

This manuscript is organized as follows: In section II, the general formalism of thermodynamic geometry considered in this work will be discussed and the key ingredients of the theoretical equation of state for the SW system will be presented. Section III is devoted to present the main results of this work about the $R$-Widom line, coexistence curves obtained with $R$-crossing method and spinodal curves for different ranges of the SW interaction. The final section is dedicated to present the conclusions of this work.

\section{Thermodynamic Geometry of the Square well fluid}

In this work, the formalism of Thermodynamic Geometry developed by Ruppeiner in \cite{Ruppeiner1979} is considered. Its origin lies in thermodynamic fluctuation theory developed by Landau \cite{Landau1937}. In this framework, the metric is related with fluctuation probability. As mentioned above, Ruppeiner metric can be obtained by calculating the negative Hessian metric of entropy, from which geometric objects of the square well fluid are then calculated. Considering the Helmholtz free energy representation, the corresponding thermodynamic metric for this potential in its matrix form is, 
\begin{equation} \label{TG:metric}
[g_{ij}] = \frac{1}{k_B T}\ \left(
\begin{array}{cc}
- \partial_T^2 f    &  0\\
0     & \partial_\rho^2 f
\end{array} \right)\,,
\end{equation}
where $f=A/V$ is the Helmholtz free energy per unit volume, $k_{_B}$ is Boltzmann's constant, $T$ stands for the absolute temperature and $\rho= N/V$ is the particle number density. It can be noticed in \cref{TG:metric}, that the corresponding thermodynamic metric for the Helmholtz free energy is diagonal for two-dimensional thermodynamic spaces, which considerably simplify the necessary calculations to obtain the required geometric objects for the analysis of the SW fluid. This is not the case for systems in higher dimensional, such as multi-component systems, see for instance \cite{Jaramillo2020,Ruppeiner2020}. \\ 
From expression given in \cref{TG:metric}, each non-vanishing component of thermodynamic metric can be expressed in terms of derivatives of reduced Helmholtz free energy $a\left(\rho^{*},T^{*}\right) = A/Nk_B T$ as a function of reduced units. 
In this work two kinds of reduced units are going to be used. The first ones are the reduced units using the interaction parameters: $T^* \equiv k_B T/\epsilon$, where $\epsilon$ stands for the potential depth;  reduced density $\rho^* \equiv \rho\sigma^3$; reduced pressure $P^* = P\sigma^3/\epsilon$. The second kind of reduced units are the ones scaled with the critical parameters: $T^*_{cr} \equiv T/T_{cr}$; $P^*_{cr} \equiv P/P_{cr}$; $\eta^*_{cr} \equiv \eta/\eta_{cr}$, where the generic quantity $A_{cr}$ represents the respective value of the critical variable. These last units will be referred as scaled units.

In terms of reduced units with respect to potential's parameters, the non-zero components  of the metric tensor are given by
\begin{align}  \label{TG:components}
g_{\text{\tiny TT}} &= \frac{1}{ \sigma^3}\left( - \rho^{*} \frac{\partial^2 a}{\partial T^{*2}} - \frac{2 \rho^{*}}{T^{*}} \frac{\partial a}{\partial T^{*}}  \right)  = \frac{1}{\sigma^{3}} G_{_{TT}}(\rho^{*},T^{*})\,, \nonumber \\
g_{\rho\rho} &= \sigma^3 \left( \rho^{*} \frac{\partial^2 a}{\partial \rho^{*2}} + 2 \frac{\partial a}{\partial \rho^{*}}  \right) = \sigma^{3}G_{\rho \rho}(\rho^{*},T^{*})\,;
\end{align}
where $G_{i i} ~(i= T,\rho)$ stand for the implicit functions defined in terms of $T^*$ and $\rho^*$, which are used in order to write scalar curvature $R$ related to metric \cref{TG:metric}  by means of a relatively simple analytic expression \cite{Ruppeiner2012},
\begin{align} \label{TG:scalar}
R = & -\frac{1}{\sqrt{g}} \frac{\partial}{\partial T}\left(  \frac{g_{_{T} \rho}}{g_{_{TT}}\sqrt{g}} \frac{\partial g_{_{TT}}}{\partial \rho} - \frac{1}{\sqrt{g}} \frac{\partial g_{\rho \rho}}{\partial T} \right)  \\ \nonumber
& - \frac{1}{\sqrt{g}} \frac{\partial}{\partial \rho}\left(  \frac{2}{\sqrt{g}} \frac{\partial g_{_{T}\rho}}{\partial T} - \frac{1}{\sqrt{g}} \frac{\partial g_{_{TT}}}{\partial \rho} - \frac{g_{_{T}\rho}}{g_{_{TT}}\sqrt{g}} \frac{\partial g_{_{TT}}}{\partial T} \right)\, ,
\end{align}
with $g$ is the metric determinant. Writing the reduced scalar curvature $R^* = R/\sigma^3$ in terms of each $G_{ii}$ as,
\begin{align} \label{TG:redscalar}
R^{*} = &  \frac{1}{\sqrt{g}} \frac{\partial}{\partial T^{*}}\left( \frac{1}{\sqrt{g}} \frac{\partial G_{\rho \rho}}{\partial T^{*}} \right) + \frac{1}{\sqrt{g}} \frac{\partial}{\partial \rho^{*}} \left(  \frac{1}{\sqrt{g}} \frac{\partial G_{_{TT}}}{\partial \rho^{*}} \right)\,.
\end{align}

\begin{figure}
\includegraphics[width=\linewidth]{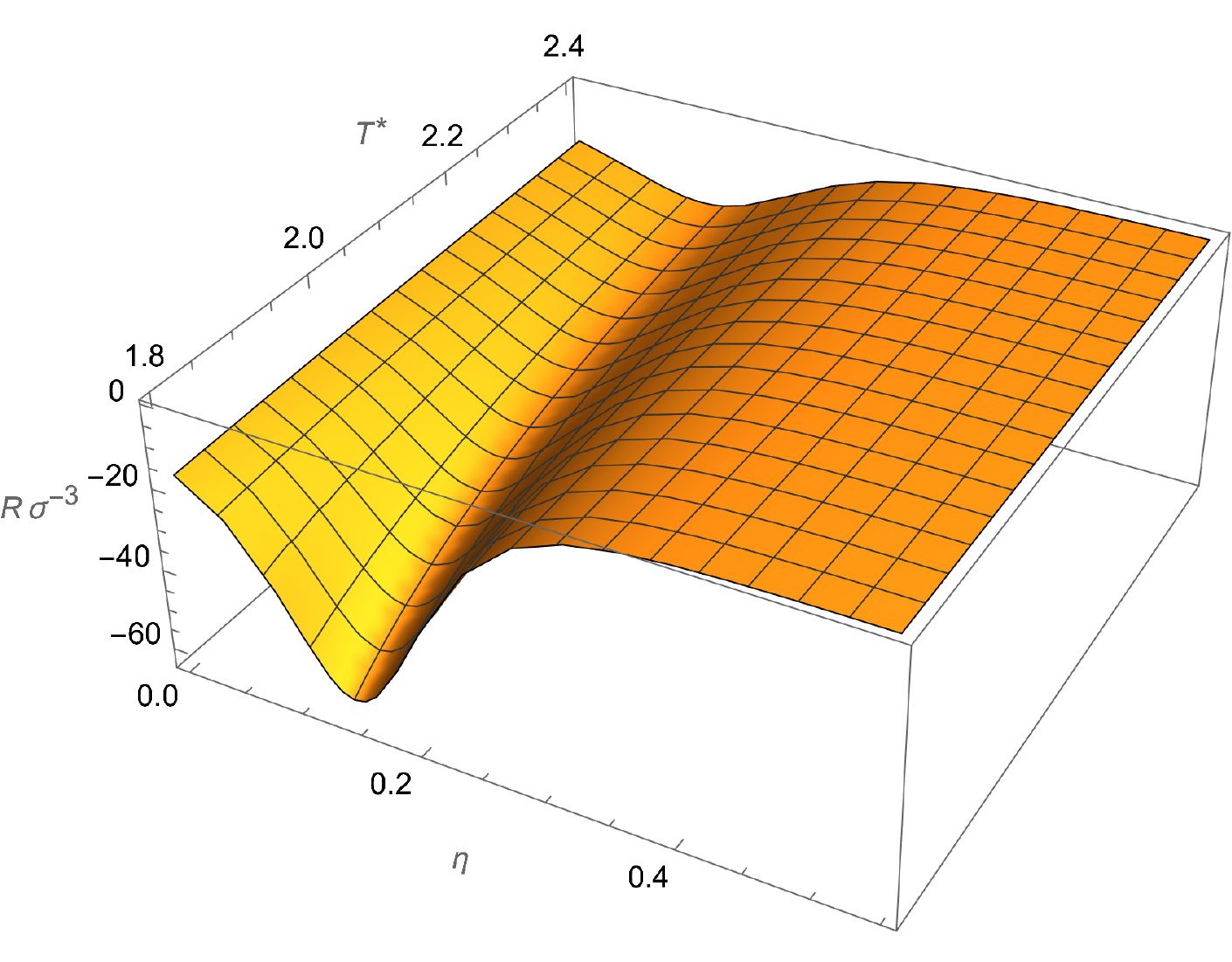}
\caption{Behavior of the reduced curvature varying reduced temperature $T^{*}$ and packing fraction $\eta$ for $\lambda^* = 1.5$ in the supercritical region. Negative curvature is related with a predominant attractive interaction.}  
\label{1}
\end{figure}

\begin{figure}
			\includegraphics[width=\linewidth]{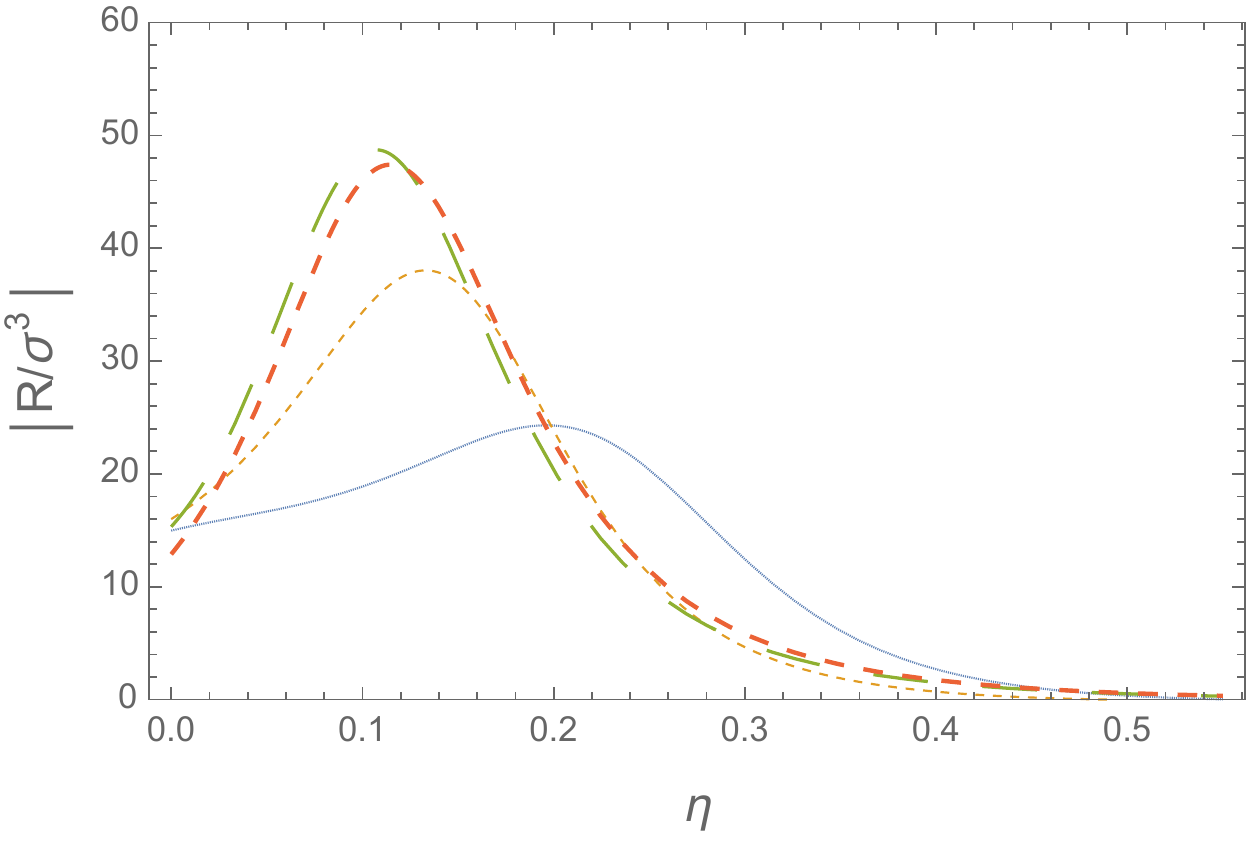}
		\caption{Isotherms of the absolute value of the reduced curvature for different values of $\lambda^*$ evaluated at 1.5 times their critical temperature; a) $\lambda^* = 1.25$ and $T^{*} = 1.254$ (continuous line) b) $\lambda^*=1.5$ and $T^{*}=1.994$ (short dashed line) c) $\lambda^* = 2$ and $T^{*}=4.237$ (medium dashed line) d) $\lambda^*=3$ and $T^{*} = 15.493$ (large dashed line).}
	\label{2}
	\end{figure}

In the thermodynamic geometry literature, it is customary to write this scalar in terms of the variables $(T,\rho)$ as  we have done so far, however,  it will be useful to rewrite these expressions in terms of the packing fraction $\eta =(\pi/6)\sigma^3 \rho$ since this variable is usually used in the construction of equations of state in the SAFT-VR formulation which is the one considered in this work for the SW fluid \cite{Gil-Villegas1997,Patel2005}. In \cref{1}, as an example, a 3D representation of the scalar curvature for the $\lambda^*=1.5$ SW fluid is given, showing the characteristic valley on the scalar curvature surface. Figure \ref{2} provides several isotherms of the scalar curvature for selected $\lambda^*$ values. It shows how the position of the extrema, as a function of $\eta$, shifts left and up as range increases. Both figures are plotted in the supercritical region. 

Using the scalar curvature $R^{*}$ given in \cref{TG:redscalar} it is possible to reproduce, to a certain extent, the coexistence curves and following Ruppeiner interpretation on the sign of scalar curvature, a negative sign is an indication of a dominant attractive behavior of the considered interaction \cite{Sarkar2012,May2013,Jaramillo2019}. Another important feature of the thermodynamic scalar curvature is the close resemblance in its behavior to response functions, whose extreme values at a given temperature in the $(P,T)$ plane can be related to the Widom line \cite{Sarkar2012,May2013}. In the following section, an analysis of thermodynamic properties of SW fluid is performed; results are compared with different standard thermodynamic methods. Before this analysis, a brief discussion on the selected equation of state for the SW fluid and its features is presented. 

\subsection{SW Helmholtz free energy}

The thermodynamic geometry formalism requires a thermodynamic potential in order to obtain the metric tensor of the associate thermodynamic space for a given system. In this subsection, a short revision of the particular expression for the Helmholtz free energy considered in this work, is provided. For a more detailed and extensive discussion regarding this expression and how it is obtained via the SAFT-VR formulation, see reference \cite{Patel2005}.\\
In a second order thermodynamic perturbation theory, the reduced Helmholtz free energy is given as an expansion in the following way
\begin{equation}
\frac{A}{Nk_{_{B}}T} = \frac{A^{\text{ideal}}}{Nk_{_{B}}T} + \frac{A^{\text{HS}}}{Nk_{_{B}}T} + \frac{A_1}{Nk_{_{B}}T} + \frac{A_2}{Nk_{_{B}}T},    
\end{equation}
where $N$ is the particle number and $T$ is the absolute temperature. $A^{HS}$ is the free energy of the reference hard-sphere system, which in the Carnahan-Starling approximation is expressed as
\begin{equation}
\frac{A^{\text{HS}}}{Nk_{_{B}}T} = \frac{4\eta -3\eta^2}{(1 -\eta)^2},
\end{equation}
with $\eta$ the packing fraction. The first perturbation term is given by,
\begin{equation}
\frac{A_1}{Nk_{_{B}}T} = - 4\left(\frac{\epsilon}{kT}\right) (\lambda^{*3} -1) \eta g^{\text{HS}}(1;\eta_{\text{eff}})\,,
\end{equation}
where $g^{\text{HS}}(1;\eta_{\text{eff}})$ is the contact radial distribution function 
\begin{equation}
g^{\text{HS}}(1;\eta_{\text{eff}})= \frac{ 1- \eta_{\text{eff}}/2}{(1 -\eta_{\text{eff}})^3},
\end{equation}
evaluated at an effective packing fraction $\eta_{\text{eff}}$ which is parameterized using a Pad\'e approximation
\begin{equation}
\eta_{\text{eff}} = \frac{c_1 \eta + c_2 \eta^2}{(1 +c_3 \eta)^3} ,   
\end{equation}
\begin{eqnarray}
c_1&=&-\frac{3.1649}{\lambda^*}+\frac{13.3501}{\lambda^{*2}}-\frac{14.8057}{\lambda^{*3}}+\frac{5.7029}{\lambda^{*4}},\\\nonumber
c_2&=&\frac{43.0042}{\lambda^*}-\frac{191.6623}{\lambda^{*2}}+\frac{273.8968}{\lambda^{*3}}-\frac{128.9334}{\lambda^{*4}},\\\nonumber
c_3&=&\frac{65.0419}{\lambda^*}-\frac{266.4627}{\lambda^{*2}}+\frac{361.0431}{\lambda^{*3}}-\frac{162.6996}{\lambda^{*4}}.\\\nonumber
\end{eqnarray}
The expansion in inverse powers of $\lambda^{*}$ for the $c_i$ coefficients, ensure that $\eta_{\text{eff}} \rightarrow 0$ for $\lambda^* \rightarrow \infty$, from which, the desired behavior in the mean-field limit: $g^{\text{HS}} (1; \eta_{\text{eff}}) \rightarrow 1$, is recovered, and $g^{\text{HS}}$ denoted the radial distribution function for a hard-sphere system.

The second order fluctuation term $A_2/Nk_{_{B}}T$ is incorporated using the local compressibility approximation. For the SW interaction can be provided in a very compact expression:
\begin{equation}
\frac{A_2}{Nk_{_{B}}T} = \frac{1}{2} \left( \frac{\epsilon}{kT}\right) K^{\text{HS}} \eta \frac{\partial}{\partial\eta}\left(\frac{A_1}{Nk_{_{B}}T}\right),   
\end{equation}
where the hard-sphere isothermal compressibility is \cite{Lafitthe2013}
\begin{equation}
    K^{\text{HS}} = \frac{(1-\eta)^4}{1+4\eta + 4\eta^2 -4\eta^3+\eta^4}.
\end{equation}
In the following table we show the critical parameters corresponding to the particular values of $\lambda^*$ chosen in this work which are obtained using the equation of state  described in this section.

\begin{center}
\begin{tabular}{|c|c|c|c|}
 \hline
$\lambda^{*}$ & $T^{*}_{cr}$ & $\eta_{{cr}}$ & $P^{*}_{cr}$ \\ \hline
$1.25$ & 0.836269  & 0.223942 & 0.164969  \\ \hline
$1.5$ & 1.32907 &  0.149937 & 0.143413  \\ \hline
$2.0$ & 2.82479 & 0.124677 & 0.232516 \\ \hline
$3.0$ & 10.3288 & 0.133652 & 0.960591 \\ \hline
\end{tabular}
\vspace{0.1cm}
\end{center}
The previous expression for the Helmholtz free energy was built in order to be accurate over a wide range of $\lambda^*$ values, specifically, it is valid for the range \mbox{$1.2 \le \lambda^* \le 3.0$}, which includes short, medium and long range interactions. Besides, another important feature of this particular equation of state is that it can be  easily extended to fluid mixtures, which is not usual for most of  proposed expressions for pure fluids.

\section{Results}

\begin{figure}[h]
		\centering
			\includegraphics[width=\linewidth]{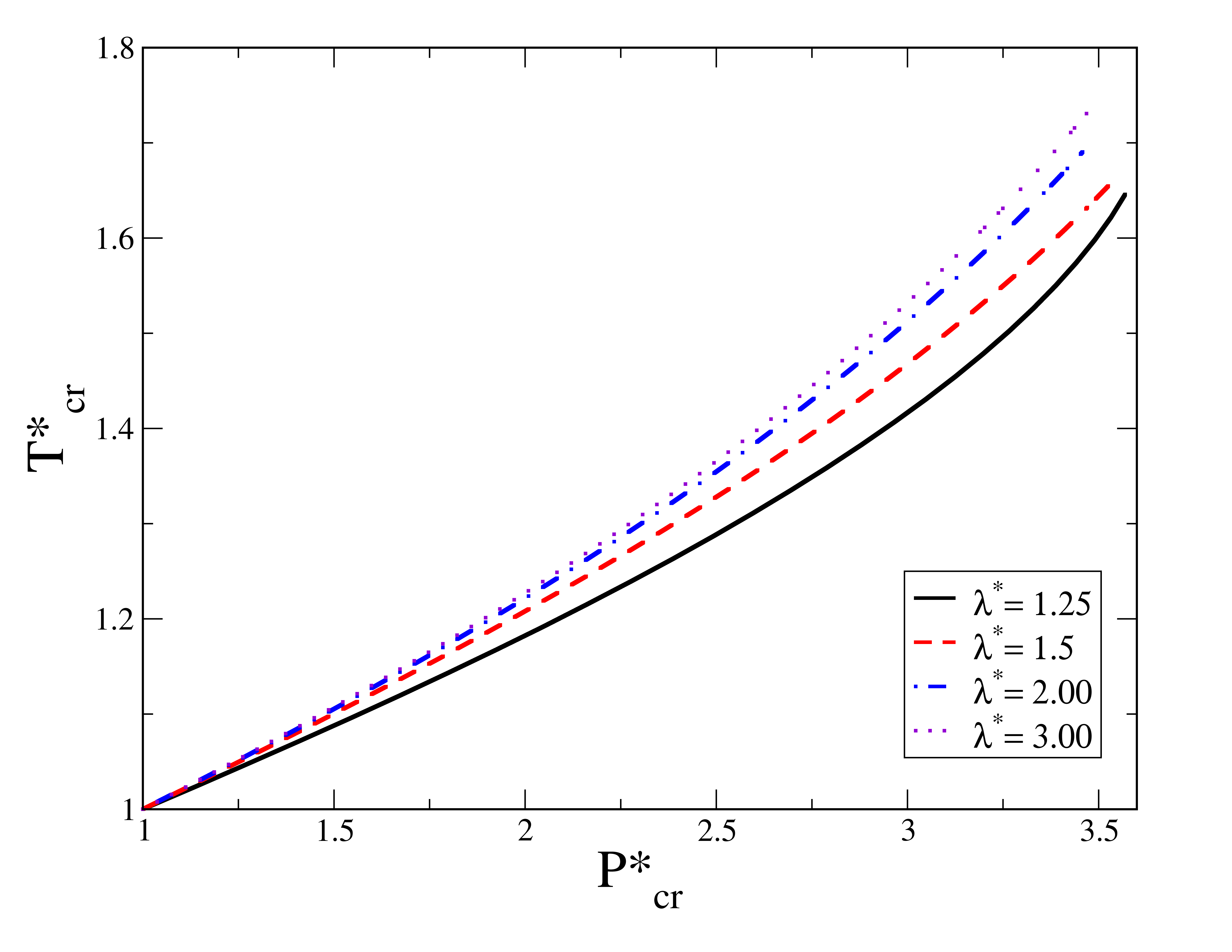}
			\caption{$R$-Widom lines in scaled units for different $\lambda^{*}$ values. It can be seen that these 
			reduced lines do not collapse onto a single curve as it might be expected if a principle of corresponding 
			states would be satisfied. There is a significant change on the slope of this lines that increases with 
			increasing $\lambda^{*}$.}\vspace{0.5mm}
			\label{3}
	\end{figure}

\begin{figure}[h]
		\centering
			\includegraphics[width=\linewidth]{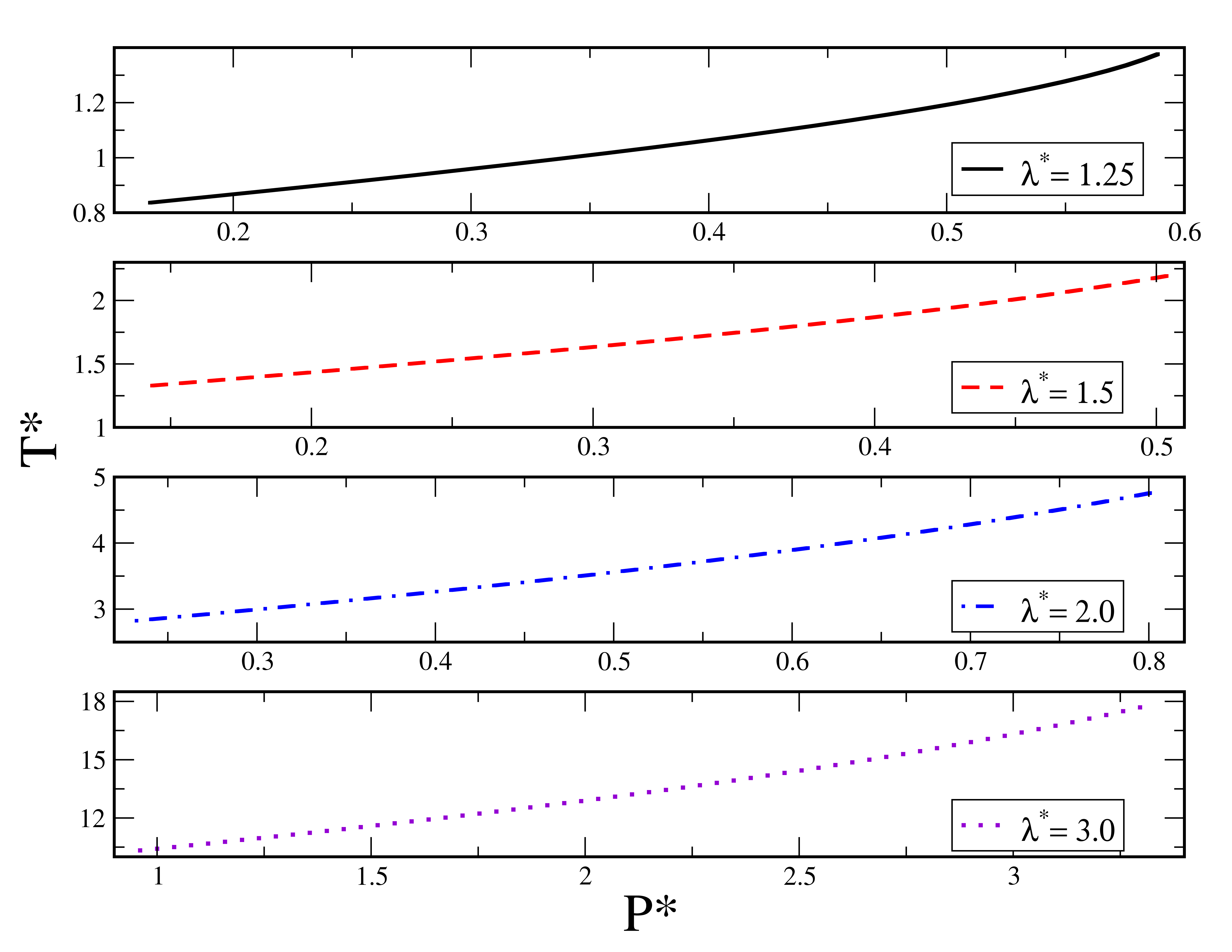}
			\caption{$R$-Widom lines in reduced (not scaled) units for different values of $\lambda^{*}$. These curves start 
			at the corresponding critical point and in the nearest region of this, the maxima of $R$ behave linearly.}\vspace{0.5mm}
			\label{4}
\end{figure}

Definition of the Widom line is not unique in the literature, for instance, it could be defined as the locus of extrema of any of the response functions 
in the supercritical region. Common definitions use extrema of heat capacity at constant pressure, isothermal compressibility and volumetric expansion coefficient \cite{May2012,Sadus2019}.\\
What will be referred as the  Widom line in this work, is the curve depicted by the extreme values of the curvature scalar $R$ in the $P^{*}-T^{*}$ plane. It starts at the critical point and in the neighborhood of this point it exhibits an almost linear behavior. In \cref{3} and \cref{4}, $R$-Widom lines are presented for different values of the range potential $\lambda^*$, in scaled and reduced units respectively (see table of critical values). 

\begin{figure}
		\centering
			\includegraphics[width=\linewidth]{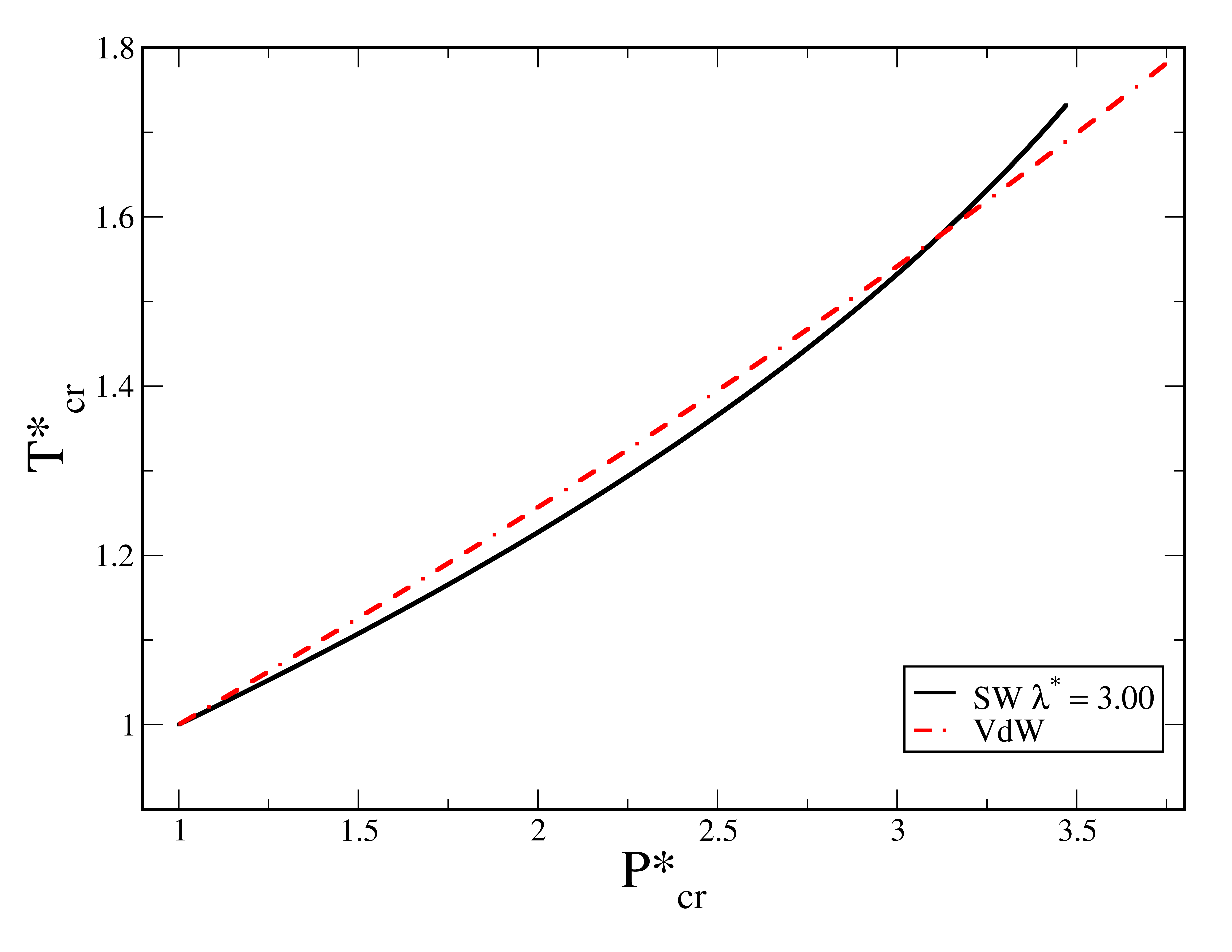}
		\caption{Widom line comparison between Square Well fluid for $\lambda^* = 3.00$ (continuous line) and van der Waals fluid (dash dot line). As the range $\lambda^{*}$ increases, the $R$-Widom line of the SW fluid gets closer to the one corresponding to that of a mean-field equation of state.}
	\label{8}
	\end{figure}

In order to display short, medium  and long-range interactions, the values $\lambda^* = 1.25, 1.50, 2.00, 3.00$ were selected. As can be seen, $R$-Widom line is sensitive to the SW range when expressed in critical reduced units and do not collapse into a single curve near the critical point. Therefore, it is not observed a kind of corresponding states law for these $R$-Widom lines. These curves move up in the $P^{*}-T^{*}$ plane as the range of the potential increases, and quickly  converge into a long-range behavior (see \cref{8}). It was expected that $R$-Widom lines do not follow a kind of corresponding states principle, since for the SW fluid, which is a non-conformal fluid, it is known that the corresponding states principle is not satisfied \cite{GilVillegas1996}.

\begin{figure}[h]
		\centering
			\includegraphics[width=\linewidth]{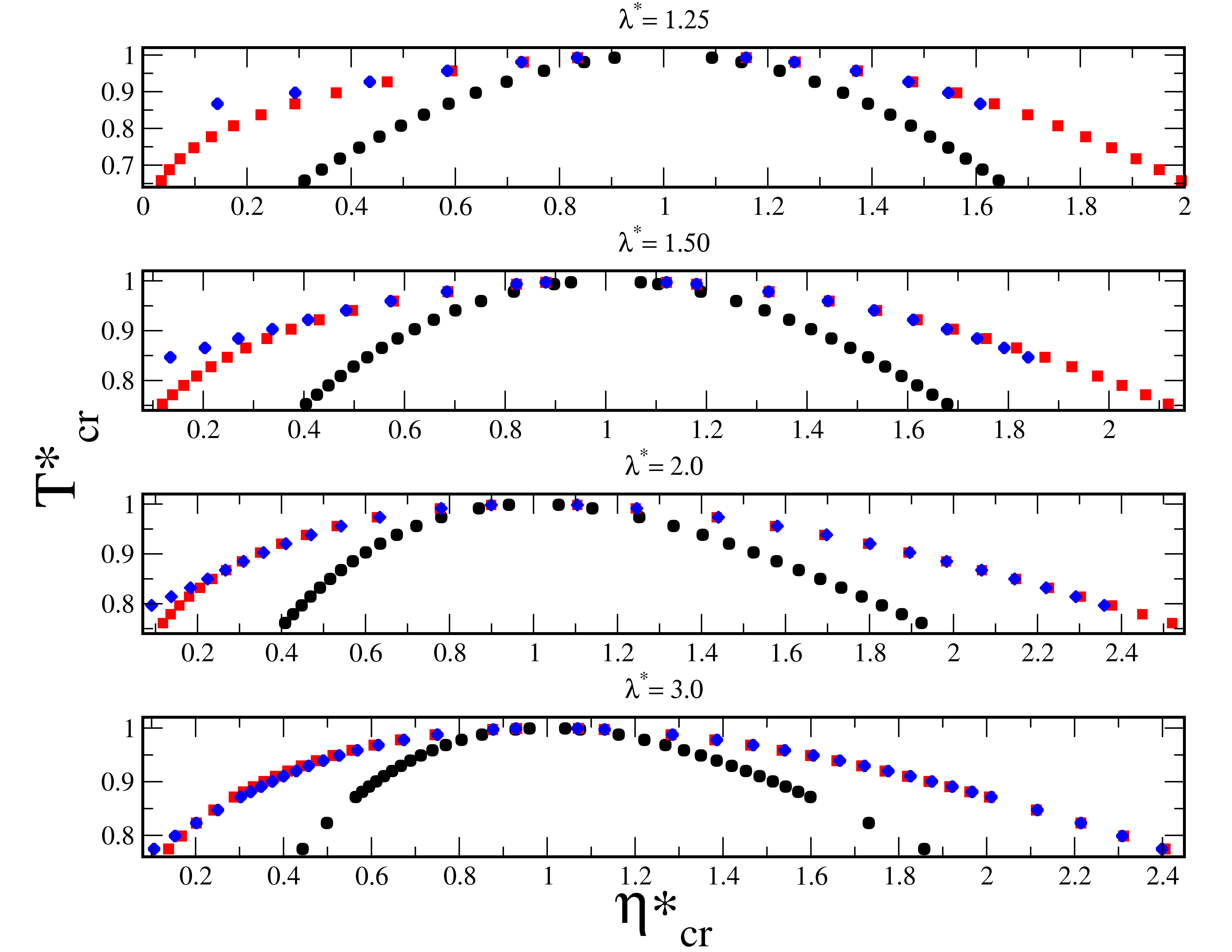}
		\caption{Reduced coexistence curves for SW fluid. Dots are spinodal points, squares stand for coexistence curves and diamonds correspond to $R$-crossing points. The geometric spinodal and geometrical curves are overlapped. The points where the curvature diverges correspond exactly to the thermodynamic spinodal curve.}
	\label{5}
	\end{figure}

\begin{figure}[h]
		\centering
			\includegraphics[width=\linewidth]{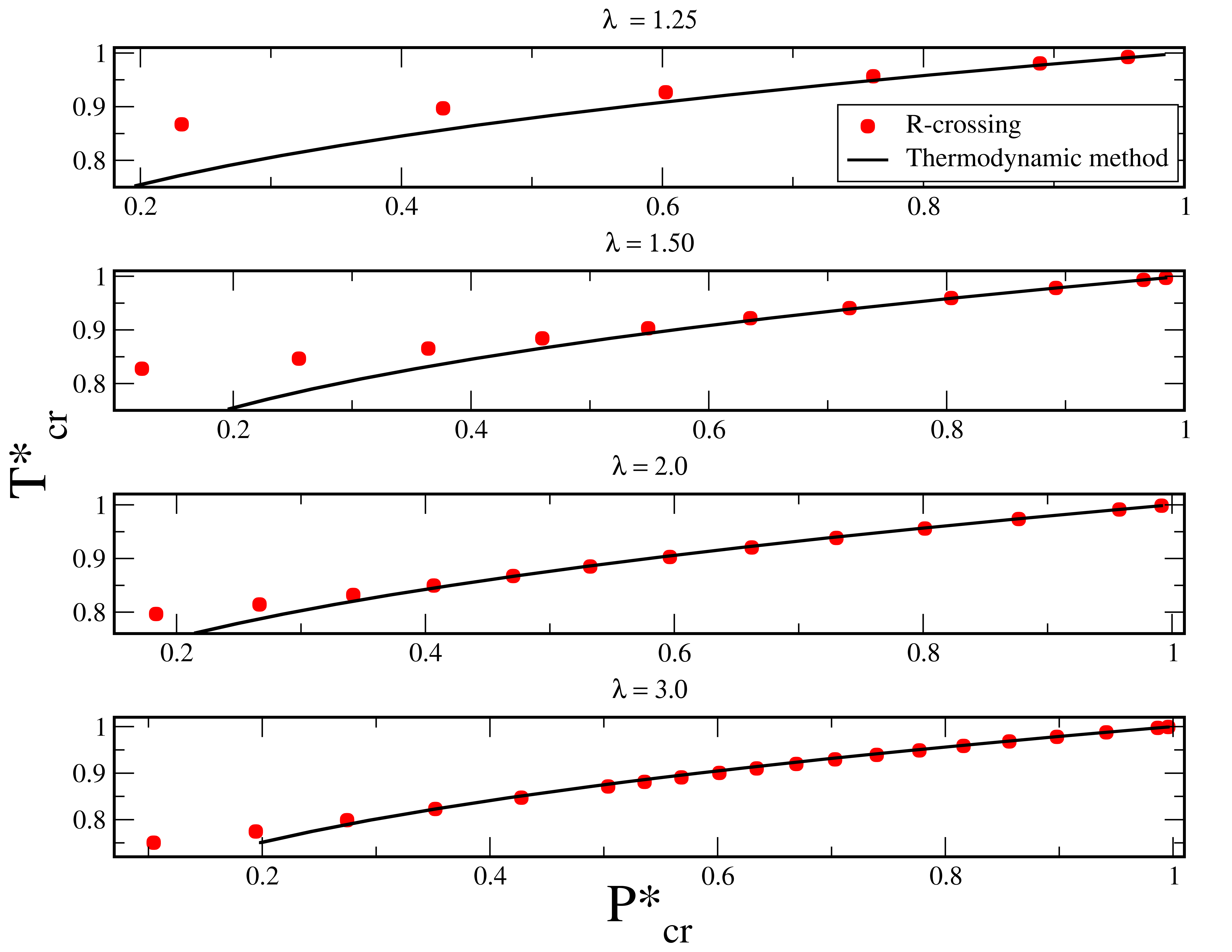}
		\caption{Saturation pressures as a function of temperature for selected $\lambda^*$ values. Continuous line represents the saturation equilibrium pressures and  red dots are the pressures where the intersections of $R$ occur.}
	\label{6}
	\end{figure}

The coexistence curves for the selected values of $\lambda^*$ using reduced critical variables are provided in \cref{5}. In order to compare, curves obtained with both methods are presented, thermodynamic standard one, where equality of pressures and chemical potentials in the two phases is solved, and the $R$-crossing method, in which the equation obtained by equality of the scalar curvature in the vapor and liquid phases is solved for a given value of temperature. It is clear that as the potential range increases, the $R$-crossing method works better for temperatures farther from the critical one. 
This can be quantitatively evaluated using the saturated pressure curves. Denoting $P(\lambda^*)$ as  the percentage at which saturated pressure curves obtained with both methods differ from each other by $5\%$ (this convention is evidently arbitrary). It is clear from the saturation pressures in \cref{6} and the curve from the $R$-crossing method in \cref{5}, that the percentage $P(\lambda^*)$ is a function of the  potential range. As it increases, the range of temperatures where the $R$-crossing method works well, also increases. In particular, for the selected ranges used in this work, $\lambda^* = 1.25, 1.5, 2.0, 3.0$, the values of the percentages  $P(\lambda^*)$ are: $6.5\%$, $8.5\%$, $15.5\%$ and $19\%$ respectively. Also in \cref{5}, spinodal curves calculated with the thermodynamic and geometric methods are shown. In the thermodynamic geometry framework, this curve correspond to all points where the curvature scalar diverges. \\
Both curves are exactly the same and consequently no difference is observed in both data sets when plotted at the same time. It is remarkable that the spinodal points can be obtained exactly by calculating the divergences of $R$ and the coexistence curve can be just partially obtained since the $R$-crossing method fails to reproduce the coexistence curve exactly at a certain point below the critical temperature.

\begin{figure}
		\centering
			\includegraphics[width=\linewidth]{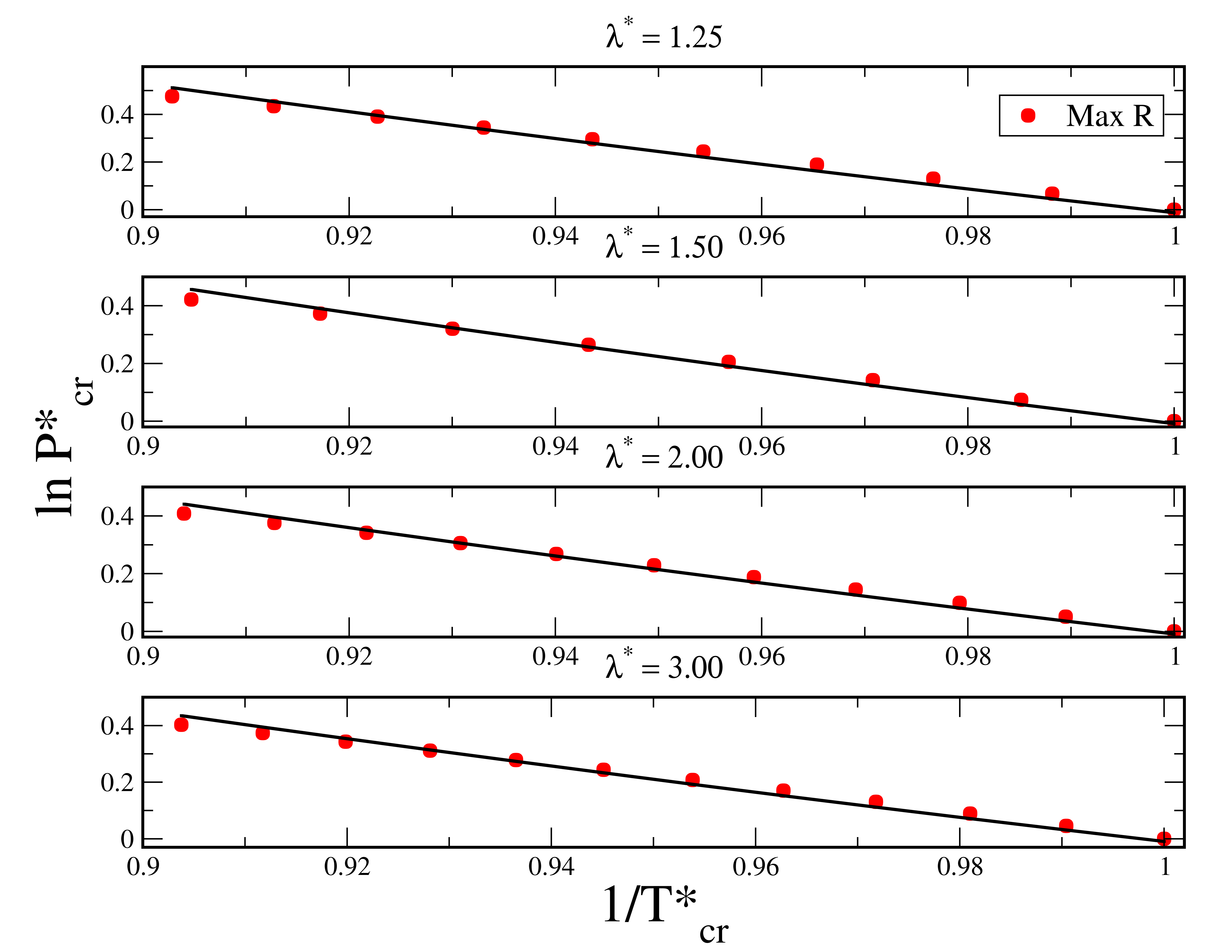}
		\caption{Logarithm of the pressures in the $R$-Widom line as a function of inverse temperature  near the critical point. Dots were obtained form the $R$-Widom line data and the continuous line represents the linear fit $\ln{P^{*}} = c + 1/T^{*}$. It is evident that near the critical point a kind of Clausius-Clapeyron is satisfied.}
	\label{7}
	\end{figure}

A very interesting behavior is observed  at the neighborhood of the critical point, a kind of Clausius-Clapeyron equation is satisfied, independently of the potential range (see \cref{7}). As is well known, an ordinary Clausius-Clapeyron equation is valid for low temperatures and pressures, and usually far below from the critical point, in such a way that the volume of the gas phase is much larger than  the liquid one. Besides, in the conventional equation, the slope of the straight line is proportional to the latent heat since,  a first-order transition occurs. From this result, it could be argued that in the supercritical region, close to  the critical point, the existence of a Clausius-Clapeyron type relationship indicates that a \textit{kind of} phase transition occurs also in this region. This is exactly the claim for the meaning of the Widom line, namely a line separating two regions above the critical point, a gas-like from another liquid-like.\\
An interesting question is: What is the meaning of the quantity playing the role of the latent heat for this curve? At this moment, possible answers to this question are not clear and more careful studies are needed. In \cref{8} a comparison between the $R$-Widom line of the $\lambda^* =3.0$ SW fluid and van der Waals one is provided. As long as the potential range increases a mean field behavior is reached.

\section{Conclusions}

In this work a hard-sphere fluid interacting trough a SW potential was studied  within the framework of Thermodynamic Geometry. Selected values for the range of the SW interaction were chosen in order to provide a global perspective of the behavior for this kind of fluid. Coexistence and spinodal curves, as well as the $R$-Widom lines were obtained  using  geometric methods. We were able to show that, unlike the coexistence curves, an exact result can be given along all the thermodynamic space and for all potential ranges for the geometric spinodal curves.\\
It was found that the coexistence curve obtained with the $R$-crossing method can be accurately reproduced below the critical temperature to a certain extent that depends on the range, being more accurate for larger ranges and for a greater percentage below the critical temperature. Additionally, the $R$-Widom lines were also given as a function of the potential range. It was also found that for the EOS used in this work, a corresponding-states principle is not found for the $R$-Widom lines.

It was shown that in the nearest neighborhood of the critical point, a kind of Clausius-Clapeyron equation is satisfied for all the interaction ranges. Since this relation is satisfied in the supercritical region, it could be argued that in the vicinity of the critical point, the existence of such relation indicates that a kind of transition occurs when crossing a boundary in a region where the Widom line is located. Whether it is just a line or a two dimensional region, is yet to date not clear \cite{Min2018}. 

\acknowledgments{J. Torres-Arenas  acknowledge support by University of Guanajuato (UG) through grant 142/2021 of Convocatoria Institucional de Investigaci\'on Cient\'ifica 2021 and by CONACYT through grant CB-2017-2018-A1-S-30736-F-2164. J. L. López-Picón was partially supported by Conacyt and UG. L.F. Escamilla-Herrera acknowledge partial support by Secretaría General, UNAM and UG.} 

\bibliographystyle{apsrev}
\bibliography{GT_Square_Well}

\end{document}